\documentstyle[11pt,moriond,epsfig]{article}

\def\be{\begin{equation}}
\def\ee{\end{equation}}
\def\bea{\begin{eqnarray}}
\def\eea{\end{eqnarray}}

\renewcommand{\[}{\begin{equation}} 
\renewcommand{\]}{\end{equation}}
\newcommand{\bef}{\begin{figure}} 
\newcommand{\ef}{\end{figure}}
\newcommand{\ie}{{\it i.e.}}

\newcommand{\llabel}[1]{\label{#1}}
\newcommand{\eq}[1]{Eq.~(\ref{#1})}

\begin{document}

\vspace{-2cm}
\noindent
{\small To appear in {\it XVIII Rencontres de Moriond: Quantum Physics
at Mesoscopic Scale} \\ \underline{Edited by
D. C. Glattli and M. Sanquer (Editions Fronti\'ers, France, 1999)}}

\vspace{1cm}
\title{INELASTIC SCATTERING AND SHOT NOISE IN DIFFUSIVE MESOSCOPIC
CONDUCTORS}

\author{YEHUDA NAVEH}

\address{Department of Physics and Astronomy, State University 
of New York \\
Stony Brook, NY 11794-3800, USA}

\maketitle

\abstracts{A short summary of the drift-diffusion-Langevin formalism
for calculating finite-frequency shot noise in diffusive conductors is
presented. Two new results are included in this presentation.  First,
we arrive at a simple (but accurate) phenomenological expression for
the semiclassical distribution function of electrons in the presence
of electron-electron scattering. Second, it is shown that in thin
samples, low-frequency shot noise may be large even if the sample
length is much larger than the electron-phonon relaxation length.}

\section{Introduction}

Non-equilibrium (``shot'') noise has been recognized in the past
decade as a valuable probe of correlations in mesoscopic electronic
systems. It was extensively studied in diverse systems ranging from
single-electron\cite{Averin 91,Birk 95} and resonant
tunneling\cite{Li 90,Iannaccone 98} devices to quantum point
contacts$^{5-9}$,
Josephson 
junctions$^{10-14}$,
and fractional 
quantum Hall effect layers$^{15-17}$
(for a review see Ref.~[18]).

In diffusive mesoscopic conductors, with elastic mean
free path $l$ much shorter than the sample length $L$, spectral
density $S_I(\omega)$ of the noise at zero frequency equals 1/3 of the
classical Schottky value $2e I$, with $I$ the average
current,\cite{Beenakker 92,Nagaev 92} a  result obtained in a
quantum-mechanical transmission 
approach\cite{Beenakker 92} and a semiclassical approach.\cite{Nagaev
92,de Jong 96} This quantum suppression of the noise is due 
to Pauli's exclusion
principle which effectively limits the phase space available for
electrons emerging from collisions with impurities, thus reducing the
randomness of 
the scattering events and enhancing correlations.\cite{Kogan 69}
However, these correlations may be strongly  affected by interactions in the
electronic system. Such interactions include the direct Coulomb
interaction manifested by 
short-range electron-electron scattering$^{23-26}$ 
and by long-range screening,$^{27-29}$
electron-phonon 
interaction,\cite{Beenakker 92,Nagaev 92,Naveh 98a} and BCS and
Andreev pairing.\cite{Naveh 99} Furthermore, the frequency dependence
of the spectral 
density may contain additional information on electronic correlations
that is not
available from the zero-frequency result.\cite{Naveh 97}

It
is not clear how to include the effect of interactions or finite frequencies in the
quantum transmission formalism for noise.\cite{Lesovik 89,Buttiker
90} However, in conductors with conductance $G \gg e^2/h$, a 
full quantum-mechanical calculation is not necessary because the
semiclassical suppression of the noise is already of the order of $e
I$, while interference effects lead at the most 
to weak localization corrections, of the order of $(e^2/h) eV \ll e
I $, with $V$ the applied voltage. Also, quantum effects due to finite frequency become appreciable only
at $\hbar \omega \sim e V$. (Weak localization and quantum frequency
corrections to the noise where studied in Refs.~[31] and [32], respectively).

Below we summarize the  ``drift-diffusion-Langevin''
formalism\cite{Naveh 97,Naveh 99a} which results in a simple recipe
for the calculation of finite-frequency noise in degenerate 
diffusive conductors, generally in the presence of interactions.
This formalism is
based on the semiclassical Boltzmann-Langevin 
equation.\cite{Kogan 69} Its range of validity is
the same as the that of the Boltzmann equation, namely,
$\lambda_F \ll l$, with $\lambda_F$ the Fermi wavelength, and $\omega
\ll eV / \hbar, 1/\tau$ with $\tau$ the elastic mean free time. We
also assume $G \gg e^2 / h$. 
We are
interested in frequencies comparable to the inverse Thouless time $1 /
\tau_T = D / L^2 \ll eV/\hbar$, with $D$ the diffusion coefficient.

\section{Theory}

According to the drift-diffusion-Langevin theory\cite{Naveh 97,Naveh 99a}
the noise spectral density as 
measured in the electrodes connecting the conductor  can be presented as
\[
\label{noise}S_I(\omega)=\frac{2G}{L}\int_{-\frac L2}^{\frac L2
}|K(x;\omega )|^2{\cal C}(x)\,dx
\]
where
\[	\llabel{corr}
{\cal C}(x) = 2 \int f_s(x, E) \left[ 1
- f_s(x, E) \right] \, dE \,
\]
 is the correlator of local
fluctuations, and $f_s(x, E)$ the symmetric part (with respect to
momentum) 
of the local steady state distribution
function of electrons at energy $E$. The response function $K(x;
\omega)$, which is 
solely responsible for 
the frequency dispersion of the noise, gives the current generated in
the electrodes by a fluctuating unit current source at $x$.  It is
dependent upon the specific geometry of the conductor, but its
integral over the sample length 
always equals 1. For example, in a ground-plane geometry it
becomes\cite{Naveh 99a} 
\[
\label{response}K(x;\omega )=\kappa \frac L2\frac{\cosh (\kappa
x)}{{\rm sinh}(\kappa L/2)},
\]
where $\kappa(\omega) =\sqrt{-i\omega /D^{\prime }}$ with $D^{\prime }=D+ G
L/C_0$ and where $C_0$ is the (dimensionless) linear capacitance between
the conductor and 
the ground plane. 

Once the electrostatic response of the system is known, the only
missing ingredient in \eq{noise} is the distribution function $f_s$. It 
is found by solving the stationary Boltzmann equation in the diffusion
approximation, 
\[\llabel{Boltzmann}
-D {d^2f_s(x, E) \over dx^2} = I(x, E),
\] 
with $I(x, E)$ the collision integral.

As an  example of using the above recipe, consider an equilibrium
situation where $f_s$ is a 
Fermi-Dirac distribution with lattice temperature $T$. Then ${\cal
C}(x) = 2T$, and 
the noise assumes the 
Johnson-Nyquist value. For typical non-equilibrium distributions ${\cal
C}(x)$ becomes much larger than the equilibrium correlator, and at $T=0$
the noise 
given by \eq{noise} is defined as the shot noise.

\section{Electron-Electron Scattering}

Eqs.~(\ref{noise}-\ref{Boltzmann}) where solved numerically in
Ref.~[26] for the case of finite electron-electron scattering,  under
the assumption of a classical 
collision integral,\cite{Gantmakher 87}
\[ \llabel{collintee}
I(\varepsilon ,\xi ) = \frac 1{\tau^{ee}_V}\int d\varepsilon'
\, \int d\omega_0
\, \left[ (1-f_s) f_s' f_s^+ (1 - f_s'^{+})  -f_s f_s' (1 - f_s^-) (1
- f_s'^{+} \right].
\]
Here $f_s = f_s(\varepsilon)$, $f_s' = f_s(\varepsilon')$, $f_s^\pm =
f_s(\varepsilon \pm \omega_0)$, $f_s'^\pm = f_s(\varepsilon' \pm
\omega_0)$, and $\varepsilon = E/eV$, $\xi = x/L$.   $\tau^{ee}_V
\propto V^{-2}$ is the electron-electron 
energy relaxation time of an electron with excess energy $eV$.  
This assumption is controversial at low voltages,\cite{Altshuler
85,Pothier 97} but is 
certainly 
valid at $eV \gg \hbar / \tau$~[36].
  
The strong dependence of the noise on the ratio
$\gamma = L/l_{ee} = L / \sqrt{D' \tau_V^{ee}}$ (at both low and high
frequencies), as found in Ref.~[26], is well within
current experimental resolution.\cite{Steinbach 96} Thus, 
shot noise measurements can be used as an independent probe of
$l_{ee}$, provided $\gamma$ is between 1 and 1000~[26]. Such
measurements should be viewed as  complimentary to 
regular magnetoresistance measurements, as they are sensitive to the
actual scattering length, which may be different from the dephasing
length measured in the latter.\cite{Altshuler 98}

However, the numerical solution of 
Eqs.~(\ref{Boltzmann},\ref{collintee}) is not simple and requires
significant computer time.
We therefore present here a
phenomenological expression which approximates the exact solution to
these equations. The expression is given by
\[ \llabel{phenomen}
f(\xi,\varepsilon) = { \left( {1 \over 2} - \xi \right) \exp \left(
- 4 G_\gamma^+ \right)
f_0 \left(\varepsilon - {1 \over 2} \right) + \left( {1
\over 2} + \xi \right) \exp \left(
- 4 G_\gamma^- \right) f_0
\left(\varepsilon + {1 \over 2} \right) + B \, {\rm tg}^{-1}(\gamma)
f_{th}(\xi, 
\varepsilon) \over \left( {1 \over 2} - \xi \right) \exp \left(
- 4 G_\gamma^+ \right)
+ \left( {1
\over 2} + \xi \right) \exp \left(
- 4  G_\gamma^- \right)  + B \,
{\rm tg}^{-1}(\gamma) }.
\]
Here 
\[	\llabel{Ggamma}
G_\gamma^\pm = \left( {1 \over 2} \pm \xi \right) \gamma^{1/2},
\]
\[
f_0 \left( \varepsilon \right) =\left[ 1 + \exp\left(
{\varepsilon / t} \right) \right]^{-1},
\]
and 
\[
f_h (\xi, \varepsilon) = \left\{ 1 + \exp\left[
{\left( \varepsilon + \xi \right) \over t_h(\xi)} \right] \right\}^{-1}
\]
with $t = T / eV$, 
$t_h(\xi)  = \sqrt{t^2 + 3\left(1 - 4 \xi^2 \right) / 4 \pi^2}$, and
$B$  a parameter to be determined.

Eq.~\ref{phenomen} has a simple physical interpretation: At $\gamma >
1$ electrons are entering the sample from both electrodes and are keeping
their original distribution up to a length scale 
$\gamma^{1/2}$~[26]. At the same time the
electrons are being thermalized inside the sample, an effect given by
the term proportional to the hot-electron distribution $f_{th}(\xi,
\varepsilon)$. $B$ serves as a  mixing constant between these two
effects. (It is straightforward to verify that \eq{phenomen} reduces
to the appropriate distributions at $\gamma \to 0$ and $\gamma \to
\infty$, and, at any $\gamma$, to the equilibrium distribution at
$\xi = \pm 1/2$.) 

Minimization of
the root-mean-square parameter
\[
\Delta = \left\{ \int_{-1/2}^{1/2} d\xi \, \int_{-\infty}^{\infty}
d\varepsilon \left[f(\xi, \varepsilon) - f_{exact}(\xi,
\varepsilon)\right]^2 \right\}^{1/2}
\]
gives 
\[
B = {5 \over 8 \pi}.
\]
(Root-mean-square minimization of the correlator ${\cal C}(\xi)$ gives
a slightly different value for $B$.
It is remarkable that the mixing parameter $B$ is almost universal, \ie,
does not depend strongly on $\gamma$.)  

Fig.~1(a) shows the exact distribution function at $\gamma=
30$ (see Ref.~[26] for similar figures at other
values of $\gamma$.) Fig.~1(b) shows the distribution function given by
Eq.~(\ref{phenomen}) at the same $\gamma$. Fig.~2(c) show the
difference between the two. As can be seen, The difference peaks at
$\delta_{\max} =6\%$ near the step-like singularity of $f$, and
rapidly falls to zero at the bulk of the sample. Fig.~2(d) shows the
values of $\Delta$ and $\delta_{\max}$ for various values of $\gamma$. As
is clear from the figure, Eq.~(\ref{phenomen}) provides a  good
approximation for $f$ only at $\gamma > 3$ (at $\gamma < 3$ the
concept of scattering length $l_{ee}$ is itself questionable). 

\begin{figure}[tb]
\vspace{-0.0cm}
{\hspace{3cm} \psfig{figure=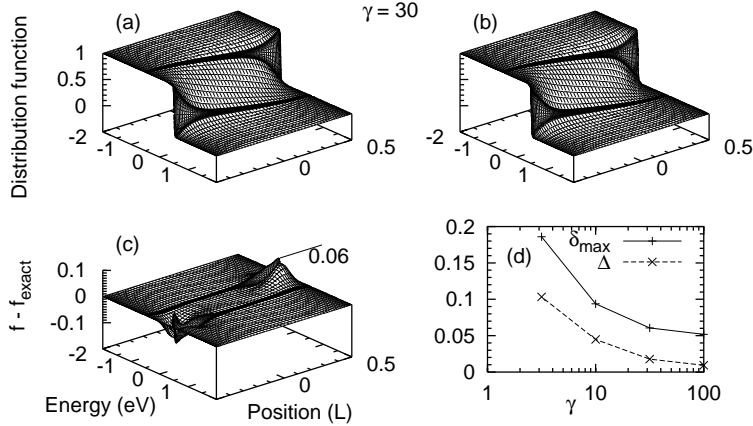,angle=-90,width=125mm}}
\vspace{-2.5cm}
{\hspace{0cm} \caption{Electron distribution function at
$\gamma=30$. (a): exact, (b): Eq.~(\ref{phenomen}), (c): the difference
between the two, (d): the maximal difference (solid line) and the
root-mean-square difference (dashed line) for various values
of $\gamma$.}}
\label{1dist}
\end{figure}

\section{Electron-Phonon Scattering}

The process of electron-electron scattering increases the noise
by virtue of adding energy to the electronic system, thus spreading
the the distribution of electrons, and increasing the correlator (\ref{corr}).
Electron-phonon scattering does the opposite: It drains energy
from the electronic system, thus reducing the spread of $f$ and
decreasing ${\cal C}$. This may suggest that when $L$ is much longer
than the electron-phonon scattering length, ${\cal C}$,
and thus the noise, vanish. As we will see below, this is indeed the
case at strictly zero frequency, but is not necessarily the case at
finite (but small) frequencies.

In order to understand this effect we first note that due to the
response function (\ref{response}), the only current fluctuations in
the conductor which are of importance in inducing noise in the
electrodes are those which are within a distance $\lambda_\omega =
1/\left| \kappa(\omega) \right|$ from the conductor-electrode
interfaces. Therefore, at high enough frequencies, the measured noise
is associated with the highly non-equilibrium distribution of
electrons near the edges of the conductor, and not necessarily with
the nearly-equilibrium distribution at the bulk of the sample. This
simple argument means that whenever $\lambda_\omega$ is smaller than
some length scale $l_S$ (which gives the spatial extent of
non-equilibrium electrons in the conductor), the shot noise value
should remain large even with increasing $L$.

To allow for a quantitative description of the above effect,
Eq.~(\ref{Boltzmann}) should be solved with the electron-phonon collision 
integral. This was done in Ref.~[30], where
it was shown 
that the width of the layer in which the electron distribution is far
from equilibrium is $l_S \approx \sqrt{L l_{ep}}$, with  $l_{ep}$ the
inelastic scattering 
length of an electron due to emission of a phonon of energy
$eV$. Therefore, one should expect 
large shot noise if 
$\lambda_\omega \approx \sqrt{D' / \omega} < \sqrt{L l_{ep}}$, or $L >
L_0(\omega)$ with 
\[	\llabel{L0}
L_0(\omega) = {D' \over l_{ep} \omega}.
\]

In what follows we would be interested in relatively long
samples and low frequencies. Therefore we assume here that $l_{ee} \ll
L, \lambda_\omega$. Having numerical results for $f$ (and
thus for ${\cal C}$) in this
situation, 
we can find the 
noise spectral density by combining Equations
(\ref{noise}) and (\ref{response}). 

Results for the noise spectral density $S_I(\omega)$ are presented in
Fig.~2 for a specific set of experimental parameters.
The upper curves in the figure show the total noise. The lower
curves show, on the same scale, the thermal noise. Since the latter is
smaller by at least an order of magnitude than the 
former, the upper curves actually depict the  shot noise. 

\begin{figure}[tb]
\vspace{0.0cm}
{\hspace{3cm} \psfig{figure=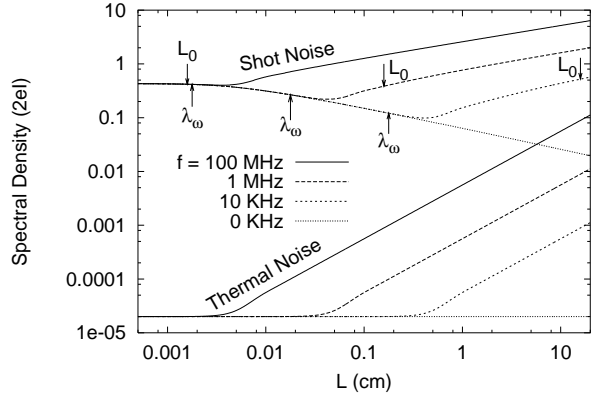,angle=-90,width=80mm}}
\vspace{0cm}
{\hspace{0cm} \caption{Noise spectral density as a function of sample's length. Lower
curves show the thermal noise and upper curves the full noise which is
dominated by the shot noise. Material parameters are $l_{ep} =
10^{-3}$ cm and $D' = 1000$ cm$^2$/s. Temperature-to-voltage ratio
is $T/eV = 10^{-5}$. Arrows indicate the positions of $\lambda_\omega$
and $L_0$ for each of the depicted frequencies (at zero frequency
these lengths tend to infinity).}}
\label{1noise}
\end{figure}

The physical discussion presented above is fully supported by the
results shown in Fig.~2. One sees that at each of the three
frequencies depicted, the noise initially decreases with $L$ up to $L
\approx \lambda_\omega$, whereupon it increases, and reaches its
mesoscopic value again at $L \approx L_0(\omega)$. The initial
decrease of the noise with increasing $L$ is due to the electrons
being increasingly thermalized in the bulk of the sample, while the
subsequent increase is due to the widening of the non-equilibrium
surface layer as $\sqrt{L l_{e p}}$, and therefore the increasing
distance from equilibrium of the noise-inducing electrons within the
layer of distance $\lambda_\omega$ from the interfaces.
As expected, at strictly zero frequency the noise reduces
monotonically to the thermal value at $L \rightarrow \infty$.

\section{Discussion}

We believe that the phenomenological expression for the distribution
function [Eq.~(\ref{phenomen})] is sufficiently accurate for all types
of calculations involving nonequilibrium electrons in diffusive
conductors, provided that $L > 3 l_{ee}$. In addition to noise
calculations, possible applications of Eq.~(\ref{phenomen}) may include 
superconducting spectroscopy experiments\cite{Pothier 97} and critical
current modulation\cite{Baselmans 99} in dirty metals between
superconductor electrodes. It is important to note that because of the singular
(step-like) boundary conditions of the distribution function at the
edges of the sample, a perturbative solution of the Boltzmann equation
in the limit $\gamma \to \infty$ is not a good approximation of
$f(x,E)$ even for very large values of $\gamma$.

The collision integral (\ref{collintee}) used here to model the
electron-electron interaction is strictly valid only at $eV \gg \hbar
/ \tau$~[36]. In this limit the 
mixing parameter $B$ in Eq.~(\ref{phenomen}) is
universal and equals to $B = 5 / 8\pi$. At lower voltages
weak localization effects become important, and Eq.~(\ref{collintee})
is no longer valid. However, even though the exact form of the
collision integral is not known, it is reasonable to assume that
equation (\ref{phenomen}) may still be valid, but with an exponent in
\eq{Ggamma}  which
scales down as the exponent of the kernel in the collision
integral,\cite{Altshuler 85,Pothier 97} and with possibly a different
value of $B$.

In the presence of electron-phonon scattering, 
the unusual result of shot noise increasing with increasing sample
length is essentially due to a competition between two independent
physical processes: screening and equilibration. The importance of
screening in affecting shot noise was first discussed by Landauer in
qualitative terms\cite{Landauer 95+96}, and was later studied
quantitatively in Refs.~[27,29]. Its outcome 
is summarized by \eq{response}. Equilibration, on the other hand
is responsible for the surface layers of
non-equilibrium electrons. The fact that the width $l_S$ of these layers
grows with $L$ is readily understood:\cite{Naveh 98a} since the
electron-phonon relaxation time decreases strongly with the energy of
the emitted phonon, at large $L$, when the electric field in the
conductor is small, an electron entering the sample from the electrode
must diffuse elastically for a long distance before being able to emit
a phonon.

Parameters chosen in obtaining Fig.~2 correspond to typical
experimental setups.$^{37,41-43}$
At frequencies higher than 1 MHz one sees that shot noise
remains of the order of $2 e I$ for any length of the conductor.
Moreover, even 'zero frequency' experiments are invariably performed
at an actual frequency of 10 KHz or higher, and should therefore
reveal large shot noise when the sample is long. While $D = 1000$
cm$^2$/s is quite realistic, 
the electrostatic term in $D'$, $G L/ C_0 \approx D (t d /
\Lambda_0^2)$, dominates if the thickness $t$ of the conductor or its
distance $d$ from the ground plane are larger than the static
screening length $\Lambda_0$. Thus, the results shown in  Fig.~1 are
of particular importance when the conductor is very thin, possibly a
two-dimensional electron gas. In addition,
for $L_0$ to be reasonably small $l_{e p}$ must be large. To
maintain $l_{ep} = 10^{-3}$ cm, $V$ cannot be larger than about
100 mV. It is therefore likely that in an actual situation $T/eV$
would not be smaller than $10^{-3}$. Then, at large $L$ and $\omega$,
the thermal noise may be as large as the shot noise.

The response function (\ref{response}), and thus the results for the
noise shown in Fig.~2, are not necessarily valid for geometries
different from the one studied here. In particular, the question of
whether any specific geometry exhibits shot noise when the conductor
is long enough reduces to the question whether finite-frequency
fluctuations in the bulk of the conductor are sufficiently screened as
to not induce current in the electrodes. Theoretically, a detailed
answer to this question may involve difficult solutions of the Poisson
equation. However, in a charged Fermi system finite-frequency currents
are known to be screened beyond some typical length scale
$\lambda_\omega'$ which does not depend on $L$~[44]. On the
other hand, the 'hot-electron' length scale $l_S=\sqrt{L l_{ep}}$
is independent of the geometry. Therefore, it is argued that
in sufficiently long samples of an arbitrary geometry $l_S$ is larger
than $\lambda_\omega'$, so the only important sources of noise are
from the non-equilibrium regions near the electrodes. Following the
physical discussion above implies that the
qualitative features of the results presented in Fig~2 may be of a
general nature.

\section{Conclusions}

The ``drift-diffusion-Langevin'' formalism for finite-frequency shot
noise in the presence of interactions was summarized. A
phenomenological equation for the distribution function in a mesoscopic
sample with $l_{ee} \leq L \ll l_{e p}$ was presented.  In the presence
of phonon relaxation, $l_{ee} \ll 
l_{ep} \leq L$,  and for thin conductors, 
it was shown that shot noise may be large even at $L \gg l_{ep}$. 
For example, for a two-dimensional electron gas near a ground plane,
shot noise at $\omega = 100$ MHz is of the order of 
the Schottky value for any length of the conductor.

\section*{Acknowledgments}
Useful discussions with D. V. Averin,
K. K. Likharev, and D. Menashe are gratefully acknowledged. The work
was supported in 
part by DOE's Grant 
\#DE-FG02-95ER14575.

\end{document}